# Intelligent Routing to Enhance Energy Consumption in Wireless Sensor Network: A survey

*Yasameen Sajid Razooqi[1], Muntasir Al-Asfoor [1]

[1]Computer Science Department, College of Computer Science and IT ,University of Al-Qadisiyah, Iraq

[1]*`yasameen.sajid@qu.edu.iq`
`muntasir.al-asfoor@qu.edu.iq`
* Corresponding Author

**Abstract.** Nowadays, the network and the Internet applications have gained a substantial importance in the digital world, due to the great impact which it provides for health and community services. Among the most important services that have been provided are smart devices and vital factor measurement devices for patients, whether in a hospital or outside the hospital. Furthermore, sensors that collect medical data or measurements of temperature and humidity, in various critical environments. The proper types of network that may be used in such difficult environment are Wireless Sensor Networks, that used to sense and process data. Additionally, the Wireless Sensor Networks have been used in the environment of Internet of Things and smart cities in the general services and health fields. All these reasons have made researchers focus on Wireless Sensor Networks and addressing the challenges that face them. The most important challenge facing this type of network is energy consumption and increase battery life. This paper discusses the methodologies used in energy conservation in Wireless Sensor Networks, such as data reduction technology, shortest path selection and artificial intelligence algorithms used in smart routing and energy saving. Besides, we have introduced comparisons between the standard algorithms which are suggested by the researchers, to make a clear picture of the energy consumption problem and propose some effective solutions in Wireless Sensor Networks field.

## 1     Introduction

In recent years have shown a significant interest in WSNs in in academia and industry due to their small size, low price, and its ability of self-configuring and organizing [1]. Those characteristics of sensors in addition to its main function of sensing physical actions made it possible to deploy a massive number of sensors in locations difficult to be managed manually [2]. The wireless Sensor network is a type of ad hoc networks, in it, the sensed data transferred to the sink or base station in multi-hop routing method through several sensors [3]. Ad-hoc network has no specific infrastructure nor centralization as in traditional network, also it is subject to topology change in the event of node move, node death or new node joins [4].WSNs applications ranging on wide and varied fields such as monitoring environment phenomena, target tracking and intrusion observation in the military, underwater searches, forests, health care, industrial machines, monitoring big structures like ships, aircraft, nuclear reactor, etc., and other applications [1].According to data transfer requirement, applications have been classified into: (i) event-driven in which data transmission start after a certain event, (ii) time-driven sensed data transmitted to the destination in a specific time and (iii) query-driven data transmitted after some request reached to the sensor [5].The most important investment of WSNs is the integration with IoT, this integration allows WSN to connect to the internet that facilitates the role of WSN in many applications. The most known applications used this integration are smart city and e-healthcare, what happens in it is accessing sensors' sensed data by the user via mobile or PCs connected to the internet [6]. Because of the importance of critical data sent in these applications, harsh monitored area and sensor operation depending on battery, it became important to discover ways for extending WSN life span and developing methods for energy consumption management [7]. In this paper, specifics of WSNs and sensors will discuss in section 2, different methods of energy maintaining in section3, discussion of the limitations and advantages of the methods in section4, and conclusion in section 5.

## 2  Wireless Sensor Networks Structure

Wireless Sensor Networks "WSNs" consisting of a large number of small size devices with constrained capabilities named sensors deployed in some area for monitoring and measuring different physical actions, connected by wireless connections for working with each other as a network [8]. The physical actions vary according to the required application, actions like pressure, light, temperature, humidity, and others utilized in applications such as health, environment observation, industry, military, etc. [9]. Thousands to billions of sensors disseminate in a large and far geographic environment or complicated systems that cannot be reached by a human, wherefore sensors must be small, inexpensive, and self-management [10]. Besides, sensors depend on the battery as a power supply which cannot be replaced nor recharged for the same fore mentioned reasons. All these conditions led to limiting in processing, communication, storage, sensing and all other abilities [8]. The main task of the sensor is to sensing or measuring surrounding activities, then transform them into signals to transfer those signals in multi-hop communication to a base station called a sink [10]. One or more of sinks connected to the WSN for collecting sensed data from sensors for more processing, analyzing or sending to the application, end-user or cloud through the internet [11]. The sink or base station has much more capabilities than sensors in term of processing, communication, storage and power [12], figure 1, shows the WSN infrastructure and its connection with an overlay network.

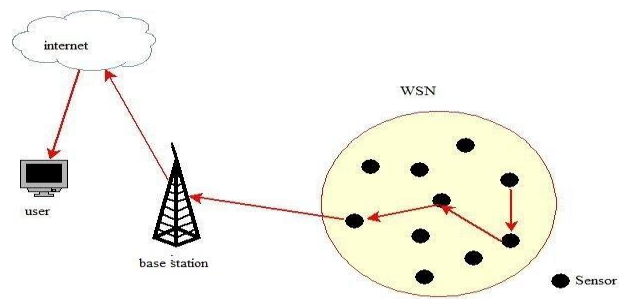

**Fig. 1.** The Basic Components of WSN

To accomplish that scenario a sensor must at least consist of three main components:
1. Sensing unit: may consist of one or more sensors for acquiring data, sensor type is application dependent [13], and may consist of analogue-to-digital converters (ADC), [10].
2. Processing unit: consist of controller and memory [10].
3. Communication unit or transceiver for sending and receive sensed data between nodes [14].

As mentioned above the battery is the source of power to the sensor which have a finite age before stop working that subsequently affecting sensor lifetime [8]. In addition to the main components, an ideal sensor has other components such as a location finding system (i.e. GPS) and mobilizer for mobility management (MM) in case of the mobile sensor, both of them is expensive and consume sensor great energy [14]. Solar cells rarely added as a source of energy in sensor [8] figure 2 shows the ideal structure of a sensor.

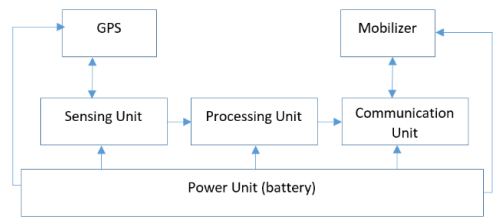

**Fig. 2.** Sensor's Structure

There are two modes for sensor distribution in the monitoring area :(i) pre-planned mode is the easier situation in it the monitoring area can be reached and manage manually which provide good coverage with less number of sensors. (ii) ad-hoc mode is very significant in the harsh or far environment or systems that cannot be reached Which requires more sensors to provide the required coverage [3]. The latter case is the most common application of WSNs with existing a non-rechargeable or replaceable battery, thus, the energy loss in some node or node death lead to all the network damage [15]. Node death and sensor mobility in case of the mobile sensor lead to a change in network topology as well new node joining or node duty cycle [8]. All mentioned above are the reasons for researchers' orientation toward energy efficiency, prolong lifetime and energy balancing of WSNs to overcome power limitation and difficulty accessing the network [16].

## 3  The Approaches to Energy Consumption Managment

The sensor's energy restriction is the major problem in WSN because it is responsible for sensor working time, which in turn affecting the network performance and lifetime [17]. Communication consumes much more energy than processing or sensing, the energy needed for transferring one bit probably enough for processing thousands of instructions [14]. Therefore, most of the methods proposed by researchers aim to perform additional computations causing in reduce the needed communication to decrease the overall wasted energy [8]. Whilst in another side some researchers proposed solutions to treat coverage holes caused by node death using mobile sensors [18], both sides targeting energy consumption management in WSNs and prolong the network lifetime as possible. The methods and techniques generally classified into sub approach as follows:

### 3.1  Intelligent Routing Protocols

Designing and developing the routing protocols in WSN is a decisive and serious process for several reasons, including energy constraints in sensor, sensor mobility and multi-hops routing due to the large distance between source and destination. Besides, the absence of centralized communication and using peer-to-peer communication between WSN nodes increases the importance of routing protocol to prevent congestion when all nodes try to communicate with each other to deliver their messages to the destination [19]. Recently, smart routing protocol development has begun taking advantage of artificial intelligent algorithms such as ant colony optimization (ACO), neural networks (NNs), fuzzy logic (FL), genetic algorithm (GA), etc. in finding best path. These algorithms improve network performance by providing adaptability to suit the change of the WSN topology, energy problem and environment complexity and changing through intelligent behavior [20]. In general, routing protocols can be classified according to the WSN architecture into:

### 3.1.1 Hierarchical Protocols

In this type, network divided into sets called clusters, each of which contains many sensors. The sensors in one cluster sense data and send them to a special node in the cluster named Cluster Head (CH). CH in turn collects data from all cluster's ordinary sensors and send them to the sink in one hop or multi-hop communication through other CHs, figure 3 illustrates clustering approach in WSN. Different techniques are used for cluster forming, electing cluster head and perhaps changing CHs periodically [21]. Clustering benefits are energy consumption balancing in-network and prevent coverage hole occurred due to nodes death near sink. The reason for that is as follows, WSN disseminates in large scale with a very huge number of sensors and various distances from the sink, therefore most sensors cannot connect to sink in one-hop. Instead, far nodes transfer data across other nodes closest to the sink in multi-hop scenario. That leads to exhaust energy in the nodes close to the sink causing node death and then network damage [19].

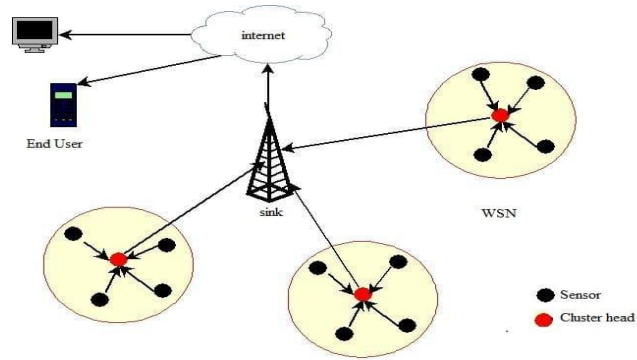

**Fig. 3.** clustering approach in WSN

In [22]; the authors introduced two methods to optimize LEACH routing protocol for enhancing the energy consumption of the nodes in WSN. The first proposed novel method enhanced the chosen method of the cluster heads (CH) which selects the appropriate CH for every cluster at each round. While the second one enhanced TDMA schedule to modify the mechanism of transmission making nodes to send nearly the same bulk of data. While, S. Govindaraj and S. N. Deepa [23] used Capsule neural network (CNN) for developing a novel learning model to perform energy optimization of WSN. The model consists of a wireless sensor layer with many sensor nodes, servers as cloud layer and sink nodes as an interconnecting layer. In Capsule Learning detecting the nodes that are appropriate to select them as cluster heads inside the cluster by identity records and also to select the forward nodes that are outside the clusters using shortest path selection thereby reducing the energy consumption. In [19] the researchers introduced a novel routing technique for clustering and routing management. They used gravitational force in addition to fuzzy rules for constructing clusters perfectly and accurately to increase the lifetime of the network. Gravitational clustering method and a gravitational routing algorithm have been proposed for finding an optimal solution for clustering and routing. Furthermore, they used a deductive inference system based on fuzzy logic for choosing suitable nodes as cluster head CH. Whereas, Muhammad Kamran Khan and his research group [24] suggested Energy-Efficient Multistage Routing Protocol EE-MRP based on clustering algorithms. In it, they divided the network area into various stages and disseminated cluster heads equally. The cluster head selection mechanism was enhanced through averting needless re-clustering by assuming a threshold in CHs selection and examination. To diminish the distance between CHs and BS, they inserted the forwarder node idea, that has extra power and its battery can be changed or recharged. All those concepts boost the network throughput and prolong the network lifetime. Greatest connectivity of nodes is the principle of Connectivity Cost Energy (CCE) Virtual Back Bone Protocol [25]. To perform that, firstly Information Table was created by Message Flooding Technique, the sink does that periodically to all nodes. Then Message Success Rate (MSR) was calculated, and if the sensor node needs to send data to the sink selects a parent node that has the highest MSR in the former hop. Secondly, Virtual backbone paths (VBPs) created by the sink through select nodes from all hops that have the highest connectivity than its neighbors, these nodes called concrete nodes. After that cluster head was chosen from the concrete nodes depending on connectivity. CH send join message to all sensors, the sensor node that receives it calculates the communication cost before joining that cluster. A threshold was put to the residual energy of the CH; CH would be changed by another concrete node if its energy was less than the threshold. Energy-efficient scalable clustering protocol (EESCP) takes into account distances between clusters and inside the cluster to produce equiponderant clusters [26]. In this protocol, a novel Particle swarm optimization technology based on Dragonfly's algorithm (DA–PSO) was utilized to choose cluster heads. In it also, a new energy-efficient fitness function was used to choose an optimal collection of CHs. EESCP works in three stages: 1- data collection in which the base station gathers details of residual energy and ID from sensor nodes. 2- cluster building to choose an optimal collection of CHs using DA–PSO by the base station, distances inside cluster reduced to lowering the communication of sending sensed data, however, distances between clusters increased to the good distribution of the CHs across the network to considerate scalability. Then the ID and role (member node or cluster head) sent to the sensor nodes, and those, in turn, choose the closest CH, CHs run TDMA scheduler to prohibit collisions inside clusters. 3- data transmission: sensors send sensed data to CHs according to TDMA scheduler, while CHs in their role transmit data to the base station. The researchers in [27] produce clustering protocol works in rounds and each round with phases aiming to achieve energy efficiency. Cluster head election is the first phase, in this phase two evolutionary algorithms was used, the first is harmony search to select a primary group of interim CHs and the second was a firefly algorithm to select and announce the final group of CHs from the candidates in the first algorithm for the present round. Cluster formation is the second phase in which normal nodes relate to CHs according to the closest distance and higher residual energy. Where the sensed data sent to the sink in the last phase after collecting data by CHs from normal nodes in the cluster using

TDMA scheduler. In firefly algorithm, they employed energy prediction, node density, and cluster compactness as parameters of a fitness function to guarantee to choose the final group of CHs those have high residual energy and expend low energy within clustering process. However, other researchers using the concept of clustering without having cluster head such as Balanced Energy Adaptive Routing (BEAR) [28].

### 3.1.2 Flat Protocols

Within these protocols, all sensors in the network send sensed data to the sink without using the cluster head in the middle for aggregating data. Sensors deliver data to the sink in either one-hop in the small networks or multi-hop connection in the wide networks. The challenge in the second case is the choice of best next relay node in the path from source to destination that consumes least possible energy or shortest path taking into account latency and other Quality of Service (QoS) aspects [10]. The researchers in [29] produced a new routing protocol called Intelligent Opportunistic Protocol (IOP) make use of artificial intelligent for choosing the probable relay node by utilizing naïve Baye's classifier to attain both energy efficiency and reliability which is the main objective within WSN. Packet Reception Ratio, Residual Energy of nodes, and Distance were the features used for getting the probability of a node to be a forwarder for next-hop. They show as a result that IOP optimizes the network lifetime, constancy, and throughput of the WSN. Whilst Jianpeng Zhang [30] decrease energy consumption and delay, balancing energy consumption and find global optimum in a neighbor choosing strategy of WSN through proposes a routing algorithm by means of optimizing the intelligent chaotic ant colony algorithm and using strategies such as neighbor chose and global optimum solution with other characteristics and conditions. In the same concern for choosing the next proper relay node, Multi-objective integer problem (MOIP) was presented in [31] to solve the trade-off between energy consumption and reliability optimization in WSNs routing. The concept of Pareto efficient solutions was used in multi-objective optimization, which introduces a series of solutions to meet halfway between the two objectives rather than the conventional thought of optimal solutions. They proposed an evolutionary algorithm based on the Non-dominated Sorting Genetic Algorithm II (NSGA-II) using the shortest paths and breadth-first search (BFS) methods to produce Pareto set in a small computational time. The researchers used real case studies to show NSGA-II's results in medium and wide size scenarios, and they prove the compromise between energy efficiency and delivery ratio. Furthermore, the performance of NSGA-II has not affected by the scalability.

### 3.2 Duty Cycle

In WSN, it is possible to sensors switch from active to a sleep state and vice versa to save power because in the sleeping state the sensor stops sensing and communication. Therefore, active mode consumes the sensor twice more times energy than sleeping, for that sensor must turn off as it can. Duty cycle is the percentage of the time the node is active for the length of the entire period [32].Optimization Duty-Cycle and Size of Forwarding Node Set (ODC-SFNS) schema was suggested in [33] to minimize delay and energy consumption. The authors supposed that the duty cycle (DC) value and the size of the forwarding node-set (SFNS) effect on latency and energy-consuming. With small DC, waiting for the forwarding node to wake up is the source of delay, but fewer collisions thus low energy consumption. In the case of large DC, more than one candidate node is active, which causes collisions leading to delay. Concerning the SFNS, the great SFNS and low duty cycle, increase the number of active forwarding nodes when data is sent that causes less latency. In contrast, great SFNS and high DC mean that many active forwarding nodes at the same time, resulting in a collision, energy consumption and delay. To overcome this tradeoff, tow strategies were adopted (1) they decreased the SFNS in dense and large DC networks, which leads to lowering one-hop delay and the number of hops necessary for the data to reach the sink, thus lowering the end-to-end delay. (2) In parse and high-DC networks, they increased the DC in the far region from the sink to reduce delay, prolong network life, and enhance energy efficiency. Finally, they combined strategy 1 and 2 to perform a comprehensive strategy.

### 3.3    Data Manipulation

The main purpose of data manipulation in WSNs is to reduce the data volume that is sent over the network in order to reduce the power spent on communications because a large amount of energy drains in communications compared to other sensor functions [14]. From the most used methods in this regard:

### 3.3.1    Data Reduction

Instead of sending all the sensed data from sensors to the sink, data reduction techniques are used to remove redundant or unnecessary data before sending. This process accomplished inside sensor nodes without neglecting data accuracy issue. Data Transmission (DaT) protocol introduced in [34] for energy management in WSN. DaT was executed locally at each sensor node periodically with two phases in every period. In the first phase, the sensed data were classified into classes using the modified K-nearest neighbor method, then the similar classes were combined into one class. While the second phase was a transmission of the data from the first phase to the sink node after they choose the best readings from the combined class rather than sending all the sensed data. The proposed protocol reduces the size of sent data which leads to decrease the energy consumption and thus prolongs the network lifetime. The same group of researchers developed their aforementioned work [35] by producing a two-layer data reduction protocol for achieving energy efficiency. In the first layer, DaT protocol, which was proposed in [34], was used in the sensor node level. Whereas in the second layer (ETDTR) protocol was implemented at the cluster head Which into it further reducing the redundant data sent from various nodes and merging them into a smaller size instead of sending all data forwarded from all nodes.

### 3.3.2 Data Compression

Data compression algorithms are used in sensors to minimize the amount of data before sending, that will subsequently minimize the communication cost as it is the biggest energy consumer in WSNs. But it is important to choose a proper compression algorithm which its execution does not consume larger energy than energy conserved in communication that leading to the same or more amount of the overall energy [36]. Adaptive Lossless Data compression (ALDC) algorithm was applied in [37] to compress sensed data in sensor before transmission using varied code choices. Two codes were developed based on Huffman coding algorithm called 2-Huffman Table ALEC and 3-Huffman Table ALEC. The proposed method was evaluated on different types of data sets considering real-time demands.

### 3.3.3 Data Fusion

Sensor readings may be affected by environmental factors like temperature, pressure, radiation or other noise, causing incorrect data. Sensor damage also gives inaccurate readings, furthermore, sensing range overlaps leads to duplicate data. Data Fusion is utilized in WSNs to remove those incorrect or duplicated data which burdens the network with needless transmission and thus an unnecessary waste of energy [38]. An improvement for LEACH clustering algorithm was performed in [39] by employing distance ratio and weighted energy, that prevents choosing the nodes with long-distance and low energy as cluster-heads. Also, they used the data fusion rate to get better performance of data transmission, this was done by cluster-heads, CHs fused data before sending to the base station.

### 3.4 Based on Mobility

In general mobility in WSNs can be achieved either by mobilizer unit, but it consumes high energy to perform sensor moving, or by a mobile entity as a vehicle, people or animals. According to application requirements or environment of the monitoring area, sink and sensors can be static or mobile. Mobility is one of the most important methods to prolong the network lifetime by reducing and balancing energy consumption and repair coverage holes [14].

### 3.4.1 Sink Mobility

Several advantages gained from sink mobility in WSN. The first is to prevent holes' formation in the network due to node death near the static sink because those nodes work in transfer other nodes data, not just their own sensed data which drains their energy faster. The second is to reduce energy consumption because the sink movement reduces the transmission distance from sensors to sink. Moreover, it enhances network performance in term of latency, throughput and connectivity [16]. The researchers in [40] discussed a solution for collecting big data from Large-Scale Wireless Sensor Network (LS-WSNs) using Mobile Data Collectors (MDCs) rather than static sinks for achieving energy efficiency because their power is renewable. They asserted that MDCs can collect data in spatially detached geographies, and reduce energy consumption in the node after divided the network into groups and clusters. For mobile data collection in WSNs, two models have proposed in this paper: data mule (MULE) for the multi-hop approach, and sensor network with mobile access point (SENMA) for the single-hop approach. These analytical approaches were employed to calculate the energy consumption in the node and to decide the optimal number of clusters for lowering energy consumption. They used a stop-and-collect protocol in which data collection takes place when the mobile collector is laid immobile at the centroids of the clusters. The proposed model showed that MULE is preferable when the number of clusters is less, and SENMA is more efficient than the MULE when the number of clusters is larger. In [41], the authors performed an enhancement to Zone-based Energy-Aware Data collection (ZEAL) routing protocol which makes efficient energy consumption and data delivery. ZEAL protocol distinguishes network into Mobile-sink node, Sub-sink node and Member node. The first gathers data from sub sink nodes by changing its location between them. The second transmit packets received from the sensor nodes to the mobile sink node. The third, sense data and transfers to the sub-sink node. E-ZEAL works in three stages: (1) pre-processing to find the optimal path for the mobile sink using the K-means clustering algorithm that minimizes energy consumption, (2) setup to enhance the choosing of sub-sink nodes that improve the data delivery, and (3) data collection. End to end delay, lifetime, remaining energy and throughput were the performance metrics used in comparing E-ZEAL with ZEAL, and E-ZEAL showed better results in all of them. Due to traffic loads in the region near to sink, sensors in this region lose their energy fast, that will eventually cause network splits and sink separation. To treat this issue, a novel routing algorithm was presented in [42], using a mobile sink that changing its location in the network, which leads to an adjustment in network energy consumption and solve sink separation by changing of its neighbors. The suggested contribution was the determining, updating and knowing the latest location of the sink by sensors to send sensed data to it, telling all nodes about that variable location consumes a lot of energy and increases latency. The proposed protocol adopted the idea of Virtual Grid-Based Infrastructure (VGB), in it, choosing some nodes called intersection nodes to save the latest location of the sink while the rest nodes could know the last location of the sink by sending a message to closest nodes in the virtual infrastructure. In addition, periodically replacing the intersection nodes if their energy diminished. Energy-efficient geographic routing algorithm was used, in which each node must know the sink location to send the sensed data to it.

### 3.4.2  Node Mobility

Data-MULE and message ferrying schemes are the most used in the mobile nodes approach. in both schemes, there existed static sensors for sensing data and mobile nodes for collecting data from those sensors and send them to the destination. These methods useful in a large scale network with sparse nodes to minimize transmission range between sensors, hence lowering communication energy [14].

Another using of mobile nodes is filling coverage holes by the movement of the near sensors to extend the network lifetime. Yijie Zhang and Mandan Liu [18] proposed a model of node positioning to resolve the coverage hole occurred by node death during the WSN working. The general idea based on a regional optimization concept, first they assessed if the node position requires to be re-optimized when some node dies, after that extract a suitable optimization region and nodes to be moved. Surrounding, redundant and mixed strategies were used to for optimization region Determining. The dynamic algorithm for regional 1Algorithm (MRDA). They suggested two objectives for the problem of node positioning optimization problem: coverage and energy consumption, with performance parameters: overall Score, the aggregate moving distance of nodes, and the rate of increasing coverage distance. They concluded that ADENSGA (a regional optimization dynamic algorithm) based MRDA (MR-ADENSGA) highly decrease moving distance for node movement which consumes great energy, improves the performance of searching and convergence velocity in WSNs with different node density. In addition, MR-ADENSGA achieved a result better in large-scale WSNs. Whilst in [43], Neighbor Intervention by Farthest Point (NIFP) algorithm consumed extra energy in moving to cover the hole occurred by node death because of energy-draining or physical jam. But this novel algorithm increases coverage with trying to minimize node moving range as it could. After calculating the size of the hole, it chooses one of the one-hop nodes in the direct neighborhood of the dead node to handle the hole area in simple computations considering the multitude of neighbors, the overlapping area with neighbors, residual energy and moving distance. NIFP compared with MDL and FSHC to show that tradeoff between energy and coverage, and high performance of NIFP in coverage with accepted energy consumption.

## 4  Disscusions

All aforementioned approaches of energy conservation in WSNs were used on the ground and proved their ability to lower energy expending and network lifetime expansion. Nevertheless, each one has drawbacks or limitations for using in some application. Nodes clustering method in WSNs has several benefits due to CHs using, that facilitate network operation, especially in wide and scalable ones as its managing cluster local routing rather than traditional routing over the whole network. In addition, data collection in CHs enables them to perform data manipulation inside the CH such as data reduction. Those factors help to save network energy as much as possible. The main limitation of this mechanism is an extra workload that leads to fast death, that because of CH work as data collector of the cluster normal nodes beside his function. Cluster formation and the most proper CH electing among cluster members are considered as challenges in hierarchal protocols, at the same time they must avoid burdening the network with great extra computation. Flat routing protocols utilize multi-hop communication to transfer packets from sensors to the sink over intermediate sensors. This scenario exhausting energy of the sink closest sensors, leaving coverage holes or causes sink separation, especially in large networks. Though these protocols suitable for medium and small networks, and those have high sensors density near the sink. The challenge in this kind of protocols is a next relay node selection that leads to the destination with fewer hops, minimum energy and delay. Also, the frequent chose to some bath more than others must be avoided to prevent unbalance energy consumption. The duty cycle method provides a direct factor in an energy saving of WSNs because of the inactive state for the sensor that reserves communication and sensing energy. The common use of this model is for dense networks in which sensors deployed randomly by helicopters in very large numbers due to anticipating some sensor damage. In this case, it's useful to turn off some nodes that have a cross sensing range alternately. Besides, it's possible to use this type for adapting network topology in some applications and changing paths as needed. The duty cycle amount must be well adjusted to avoid collision and delay and hence extra energy-consuming. If the duty cycle was very small, it will increase the probability of sleepy receivers that make the sender wait to sensors' waking up causing a delay in packets delivery that in turn causing energy loss. While the high duty cycle increases the probability of the active sensors at the same time, which leads to conflict or collision hence increase the delay and energy loss. The data reduction, compression and fusion have the same goal of minimizing the amount of data that sent over the network to save energy caused by needless data transmission. Also, minimizing the data amount can reduce delays and the size of the memory needed to store data in the sink. These techniques are undesirable in some applications that required high data accuracy as they may be causes reducing data precision or losses few data. From another side, processing of these techniques should not take a lot of time or energy that is equivalent to what already saved through communications lowering. Mobile sensor or sink is a very suitable solution for energy loss in sparse or low-density networks. In this network, mobile sinks moving across the network to collect data from static sensors or mobile sensors changing their location for collecting and sensing data. In both cases, the routing distance and the number of hops will be reduced in packets sending leads to energy saving. Changing the sink location will prevent nodes death in the nearest region to the sink because of the workload resulting from multi-hop data transmission. Nodes death causes holes, can be fixed by changing mobile sensor location. All that prolongs the network lifetime due to energy consumption balancing and provide good coverage. The drawback of this way is making the sensor waiting for the mobile collector to reach their communication range before sending data that will cause a delay. Another drawback is when using mobilizer unit to perform the movement, this expensive unit consumes the sensor great energy unless reducing the moving distance as possible or uses another way such as animals or vehicles. Finally, some environmental obstacles may obstruct node or sink movement which makes this approach inappropriate for some applications as adaptive the shape of the topology of overlay network [44]. Table 1. shows a summary of the papers that have been discussed in this survey with their distinguishing characteristics.

**Table 1.** summary of the papers that have been discussed in this survey

| Authors | algorithm | protocol | objectives | Monitoring area | Number & kind of nodes | kind & number of Sink | The initial energy of each node | Energy conservation approach |
|---|---|---|---|---|---|---|---|---|
| Mohamed Elshrkawey 2017…[22] | TDMA schedules | LEACH | energy efficiency, network lifetime | 100m*100m | 100 static | One static sink | 5 J | Hierarchical Protocols |
| S. Govindaraj… 2020 [23] | CNN | — | energy efficiency, network lifetime | 1000 m*1000 m | 200 static | One static sink | 2 J | Hierarchical Protocols |
| Munuswamy Selvi … 2020 [19] | Gravitational clustering method, fuzzy rules | gravitational routing | energy efficiency, network lifetime, delay | 200 m* 200m | 20–1000 static | One static sink | 0.5 J | Hierarchical Protocols |
| Muhammad Kamran Khan 2018…[24] | ___ | EE-MRP | Throughput, network lifetime, energy efficiency, | 150 m* 150m | 100 static | One static sink | 0.5 J | Hierarchical Protocols |
| E. Golden Julie 2016 …[25] | ----- | CCE Virtual Backbone cluster-based routing protocol | reliability, delay, energy consumption | 500 m * 500 m | 50 to 200 static | One static sink | 1 J | Hierarchical Protocols |
| Harmanpreet Singh 2019…[26] | DA–PSO | EESCP | energy efficiency, scalability. | 100 * 100 m2  200 * 200 m2  300 * 300 m2 | 100 static  200  300 | One static sink | 2 J | Hierarchical Protocols |
| Anupkumar M. Bongale 2019…[27] | firefly and harmony search algorithms. | ------- | energy efficiency, coverage | 250*250 m2 | 100 static  200  300  400 | One static sink | 2 J | Hierarchical Protocols |
| NADEEM JAVAID 2017…[28] | intra _SEB , inter _SEB | BEAR | energy efficiency, network lifetime | Radius:  0.2-1 km  1-5 km | 80 static  160 | One static sink | 0.5 J  10 J | Hierarchical Protocols |
| Deep Kumar Bangotra 2020…[29] | Naïve Bayes conditional probability | IOR | reliability, energy efficiency | 500m*500 m | 100 static | One static sink | 0.5 J | Flat protocol |
| Jianpeng Zhang 2020 [30] | intelligent chaotic ant colony avlgorithm | — | energy efficiency, network lifetime, delay | — | 0-60 static | | — | Flat protocol |

| | | | | | One static sink | | |
|---|---|---|---|---|---|---|---|
| Marlon Jeske 2020 …[31] | NSGA-II BFS shortest paths | CTP | energy efficiency, reliability | 10, 000 m2 | 100 static | One static sink | 100 J | Flat protocol |
| FEIHU WANG 2019…[33] | ODC-SFNS | — | delay, energy consumption, network lifetime | Radius: 500 m | varied density, static | One static sink | — | Duty cycle |
| Rafal alhussaini 2018 …[34] | K-nearest neighbor method | DaT | energy efficiency, network lifetime, data accuracy | — | 54 static | One static sink | — | data reduction |
| Ali Kadhum Idrees 2020 …[35] | K-nearest neighbor method | DaT , ETDTR | energy efficiency, network lifetime, data accuracy | — | 47 static | One static sink | — | Hierarchical Protocols and data reduction |
| Jonathan Gana Kolo 2012.. [37] | ALDC, Huffman coding | --------- | Compression, energy efficiency | --------- | ---------- | One static sink | --- | Data compression |
| Wenliang Wu 2017…[39] | data fusion rate | LEACH | energy efficiency, network lifetime | 100m *100 m | 100 static | One static sink | 0.5 J | Hierarchical Protocols and data fusion |
| Kenneth Li-Minn Ang 2018…[40] | MULE, SENMA | Termite-hill, MCBR_Flooding | energy consumption, network lifetime | 10 *10 km2 | 10000 static | One mobile sink | — | Sink mobility |
| Aya H. Allam 2019…[41] | K-means clustering algorithm | ZEAL | data delivery, energy consumption, delay | 400 m* 200 m | 120 static | One mobile sink | 3000J | Sink mobility |
| Yijie Zhang 2020…[18] | MR-ADENSGA | LEACH | Coverage, energy efficiency | 100m*100m 140m*140m 170m*170m 200m*200m | 50 100 150 200 , mobile | One static sink | 0.5 J | Sensor mobility |
| Banafsj Khalifa 2017…[43] | NIFP | ---- | Coverage, energy efficiency | 300m*300m | 90 , mobile | One static sink | 1 – 100 J | Sensor mobility |

## 5   Conclusions

In this review paper, we have introduced the main approaches of energy consumption management in Wireless Sensor Network, and some methods used to avoid energy-wasting. Special attention was devoted to the latest methods proposed by researchers to solve energy issues in such kind of network. The survey classified those methods into smart routing, duty cycle, data manipulation and mobile sensors or sink to minimize the communications because it's wasting energy more than computation. This literature figures out the limitations and advantages of using each method and compare the latest suggested solutions of the WSN energy conservation. The choosing of the best method for lowering energy consumption is application dependent. The purpose of the network, the location of the monitoring area, the network size, environmental factors, network density, and sensor and sink kind (static or mobile) are the most the factors must be noticed in choosing the energy preservation approach.